# A Generalized approach for computing the trajectories associated with the Newtonian N Body Problem

AbuBakr Mehmood, Syed Umer Abbas Shah and Ghulam Shabbir

Faculty of Engineering Sciences,

GIK Institute of Engineering Sciences and Technology

Topi, Swabi, NWFP, Pakistan

Email: shabbir@giki.edu.pk

**Abstract:**

The Classical Newtonian problem of describing the free motions of N gravitating bodies which form an isolated system in free space has been considered. It is well known from the Poincare's Dictum that the problem is not exactly solvable. Sets of N body systems composed of masses having spherical symmetry, appropriate angular velocities (< 1 rad/s) and bounded position vectors are examined. A procedure has been developed which yields expressions approximately defining the trajectories executed by the masses.

**Introduction:**

The problem has been solved in a two dimensional plane. Sets of N body systems in which the gravitating masses possess spherically symmetric mass distributions, small angular velocities (< 1 rad/s) and bounded position vectors have been considered. Other approaches can be found in [1-5]. The following figure shows our configuration for the system of N bodies here $\vec{x}_k$ is the position vector of mass $m_k$, $\theta_k$ is the rotation angle of $\vec{x}_k$, $\hat{e}_{x_k}$ is the radial unit vector along $\vec{x}_k$ and $\hat{e}_{\theta_k}$ is a unit vector perpendicular to $\hat{e}_{x_k}$ $\forall$ $1 \leq k \leq N$. We assume that the *kth* body having mass $m_k$ and position vector $\vec{x}_k$ remains approximately in two body motion with a body of mass $M_k = \sum_{n=1}^{N} m_n$ $(n \neq k)$, placed at a point given by $\vec{x}_{M_k} = \frac{\sum_{n=1}^{N} m_n \vec{x}_n (n \neq k)}{\sum_{n=1}^{N} m_n (n \neq k)}$. The position vectors of $m_k$ and $M_k$ are assumed to remain approximately collinear for all time in accordance with this assumption. It then follows that the respective unit vectors in those directions also remain approximately collinear for all time, that is $\hat{e}_{x_k} \cdot \hat{e}_{x_{M_k}} \simeq -1$ $\forall$ $t$. Since we are free to scale each position vector along its direction, the boundedness of the position vectors ($|\vec{x}_k(t)| < \infty$ $\forall$ $1 \leq k \leq N$) can be modified as $|\vec{x}_k(t)| \ll \infty$ $\forall$ $1 \leq k \leq N$. The closed form approximations that we will find for the case of $N$ bodies will be valid for the case $|\dot{\theta}_k(t)| < 1$ rad/s & $|\vec{x}_k(t)| \ll \infty$ $\forall$ $1 \leq k \leq N$. Note that these constraints include almost all practically encountered situations, in the sense that angular velocities encountered in celestial motion are normally much less than 1 rad/s and position vectors are bounded. These conditions therefore negligibly limit the application of our results.

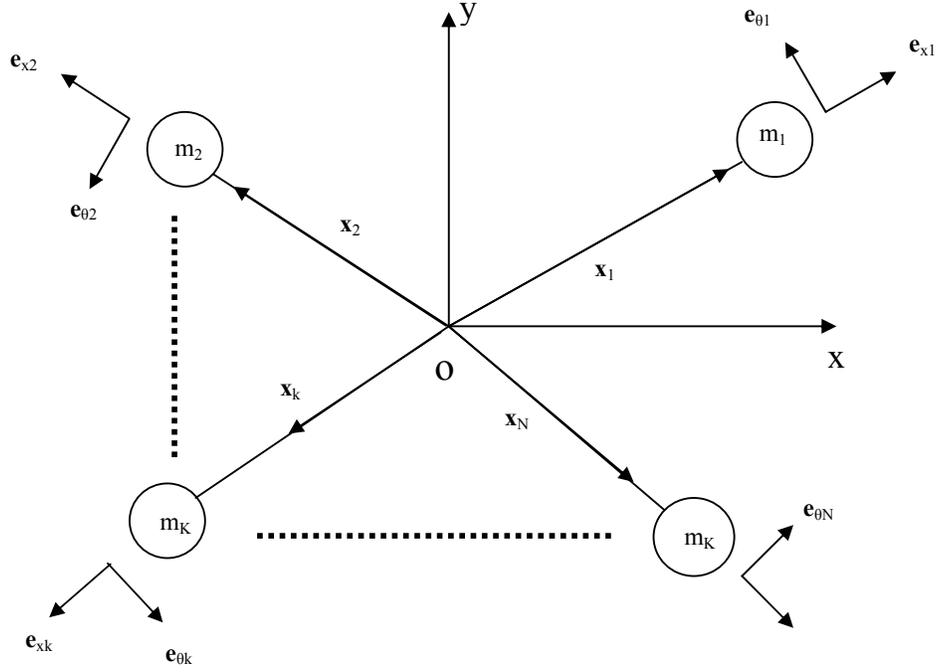

*Figure 1*

## Main Results:

Figure 2 presents the configuration of our two body approximation of the system of $N$ bodies

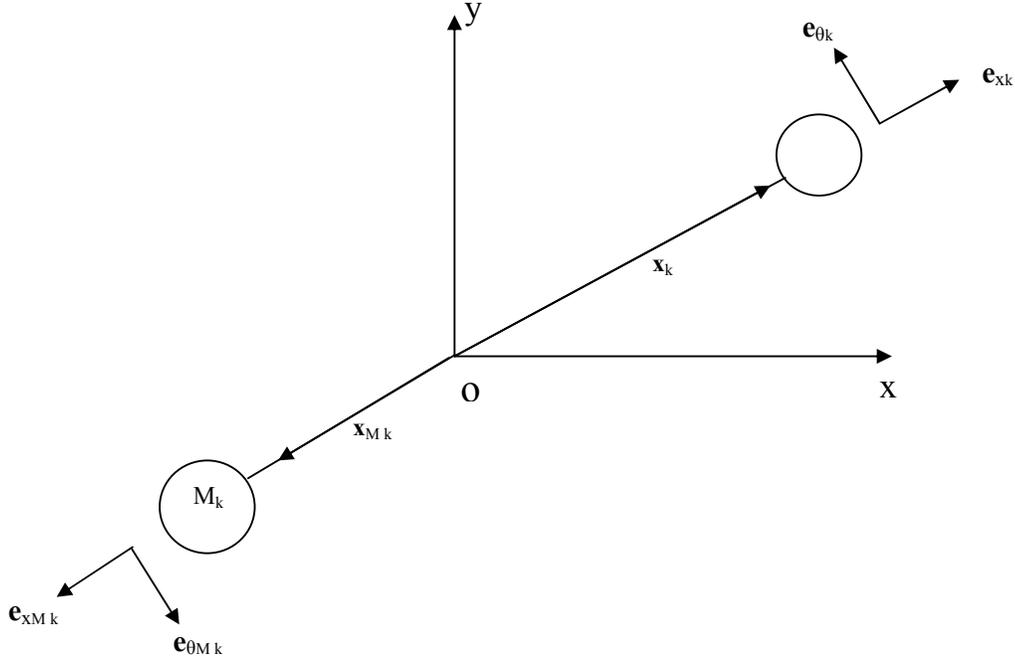

*Figure 2*

Here $\hat{e}_{x_{M_k}}$ is a unit vector in the direction of $\vec{x}_{M_k}$ and $\hat{e}_{\theta_{M_k}}$ is a unit vector perpendicular to $\hat{e}_{x_{M_k}}$. The vectors $\vec{x}_k$ and $\vec{x}_{M_k}$ are taken to be approximately collinear for all time, resulting from the two body motion analogue in figure 2. This in turn implies that $\theta_k(t) = \theta_{M_k}(t) + \pi \Rightarrow \dot{\theta}_k(t) = \dot{\theta}_{M_k}(t) \Rightarrow \ddot{\theta}_k(t) = \ddot{\theta}_{M_k}(t) \ \forall \ t$. Using Newton's Laws to model the system in figure 2 we get

$$m_k \ddot{\vec{x}}_k = -\left(\frac{G m_k M_k}{\left|\vec{x}_{M_k} - \vec{x}_k\right|^2}\right)\hat{e}_{x_k} \qquad (1)$$

$$M_k \ddot{\vec{x}}_{M_k} = -\left(\frac{G m_k M_k}{\left|\vec{x}_{M_k} - \vec{x}_k\right|^2}\right)\hat{e}_{x_{M_k}} \qquad (2)$$

where G is the universal gravitation constant. The representation of vectors $\vec{x}_k$ and $\vec{x}_{M_k}$ as $x_k(t)\hat{e}_{x_k}$ and $x_{M_k}(t)\hat{e}_{x_{M_k}}$ respectively, will find use in the scalar representation of (1) and (2) which can be shown to be

$$\ddot{x}_k - x_k \dot{\theta}_k^2 = -\left(\frac{GM_k}{\left|\vec{x}_{M_k} - \vec{x}_k\right|^2}\right) \tag{3}$$

$$x_k \ddot{\theta}_k + 2\dot{x}_k \dot{\theta}_k = 0 \tag{4}$$

$$\ddot{x}_{M_k} - x_{M_k} \dot{\theta}_{M_k}^2 = -\left(\frac{Gm_k}{\left|\vec{x}_{M_k} - \vec{x}_k\right|^2}\right) \tag{5}$$

$$x_{M_k} \ddot{\theta}_{M_k} + 2\dot{x}_{M_k} \dot{\theta}_{M_k} = 0 \tag{6}$$

The system of equations (3), (4), (5) and (6) is not exactly solvable. We recall that the problem is being solved for the case $\left|\dot{\theta}_k(t)\right| < 1$ rad/s and $x_k(t) \ll \pm\infty$ ($\forall\ 1 \leq k \leq N$). We could find an approximation to equation (3) by noting that $\dot{\theta}_k^2 \simeq 0\ \forall\ t \Rightarrow x_k \dot{\theta}_k^2 \simeq 0\ \forall\ t$ (since $\left|\dot{\theta}_k(t)\right| < 1$ rad/s and $x_k(t) \ll \pm\infty$). This implies that $\ddot{x}_k - x_k \dot{\theta}_k^2 \simeq \ddot{x}_k$ and hence the following is found to be an approximation to (3).

$$\ddot{x}_k \simeq -\frac{GM_k}{\left|\vec{x}_{M_k} - \vec{x}_k\right|^2} \tag{7}$$

It can be shown that if $|\vec{x}_k| \ll \infty\ \forall\ 1 \leq k \leq N$, then $|\vec{x}_{M_k}| \ll \infty\ \forall\ 1 \leq k \leq N$. Also, if $\left|\dot{\theta}_k(t)\right| < 1$ rad/s, then $\left|\dot{\theta}_{M_k}(t)\right| < 1$ rad/s. Using these arguments, just like we found (7) to be an approximation to (3), we can find (8) to be an approximation to (5).

$$\ddot{x}_{M_k} = -\left(\frac{Gm_k}{\left|\vec{x}_{M_k} - \vec{x}_k\right|^2}\right) \tag{8}$$

We now define the vector $\vec{x} = \vec{x}_{M_k} - \vec{x}_k$ and therefore

$$\left|\vec{x}\right|^2 = \left|\vec{x}_{M_k} - \vec{x}_k\right|^2 \tag{9}$$

Since $\vec{x}$ is a vector along the direction of $\vec{x}_{M_k}$, it follows that a unit vector along the direction of $\vec{x}$ is the same as a unit vector along the direction of $\vec{x}_{M_k}$. We define $\hat{e}_x$ to be a unit vector along the direction of $\vec{x}$. Hence it follows that $\hat{e}_x = \hat{e}_{x_{M_k}}$. Also $\vec{x}$, $\vec{x}_k$ and $\vec{x}_{M_k}$ can be represented as products of scalar time functions and respective unit vectors.

$$\vec{x} = x(t)\hat{e}_x$$

$$\vec{x}_k = x_k(t)\hat{e}_{x_k} \tag{10}$$

$$\vec{x}_{M_k} = x_{M_k}(t)\hat{e}_{x_{M_k}}$$

and

$$\hat{e}_x = \hat{e}_{x_{M_k}} = -\hat{e}_{x_k} \tag{11}$$

Using the first equation in relation (12) we can rewrite (11) as

$$\left|\vec{x}\right|^2 = \left|x\hat{e}_x\right|^2 = \left|\vec{x}_{M_k} - \vec{x}_k\right|^2 = x^2 \tag{12}$$

Rewriting (7) and (8) while making use of (12) we get

$$\ddot{x}_k = -\left(\frac{GM_k}{x^2}\right) \tag{13}$$

$$\ddot{x}_{M_k} = -\left(\frac{Gm_k}{x^2}\right) \tag{14}$$

Adding the above two result we get

$$\ddot{x}_k + \ddot{x}_{M_k} = -\left(\frac{G(m_k + M_k)}{x^2}\right) \tag{15}$$

Making use of (10) and (11) it can be shown that

$$\ddot{x}(t) = \ddot{x}_k(t) + \ddot{x}_{M_k}(t) \tag{16}$$

Substitution of equation (15) in equation (16) then yields

$$\ddot{x} = -\left(\frac{G(m_k + M_k)}{x^2}\right) \tag{17}$$

We now integrate (13) and (14) twice with respect to time to get (18) and (19) respectively.

$$x_k(t) = x_{ko} + \dot{x}_{ko}(t - t_o) - GM_k \int_{t_0}^{t}\left[\int_{t_0}^{t} x^{-2}(t)dt\right]dt \tag{18}$$

$$x_{M_k}(t) = x_{M_{k0}} + \dot{x}_{M_{ko}}(t - t_o) - Gm_k \int_{t_0}^{t}\left[\int_{t_0}^{t} x^{-2}(t)dt\right]dt \tag{19}$$

where $x_{ko} = x_k(0)$, $x_{M_{k0}} = x_{M_k}(0)$, $\dot{x}_{ko} = \dot{x}_k(0)$, $\dot{x}_{M_{ko}} = \dot{x}_{M_k}(0)$ and $t_0$ is the starting time. In both (18) and (19) the expression $\int_{t_o}^{t}\left[\int_{t_o}^{t} x^{-2}(t)dt\right]dt$ is unknown. We can integrate equation (17) twice with respect to time to find an alternative form for the above expression as

$$\int_{t_o}^{t}\left[\int_{t_o}^{t} x^{-2}(t)dt\right]dt = \left(\frac{\dot{x}_o(t-t_o)+x_0-x(t)}{G(m_k+M_k)}\right) \quad (20)$$

where $x_0 = x(0)$ and $\dot{x}_o = \dot{x}(0)$. Now substituting equation (20) into (18) we get

$$x_k(t) = x_{ko} + \dot{x}_{ko}(t-t_o) + \left(\frac{M_k}{(m_k+M_k)}\right)[x(t)-\dot{x}_o(t-t_o)-x_o] \quad (21)$$

In order to find the analytic expression for the trajectory of $m_k$ by use of equation (21), we must find $t(\theta_k)$. Therefore we rewrite (21) as

$$x_k(\theta_k) = x_{ko} - \left(\frac{M_k}{(m_k+M_k)}\right)x_0 + \left[\left(\frac{M_k}{(m_k+M_k)}\right)\dot{x}_o - \dot{x}_{ko}\right]t_o$$

$$+ \left[\dot{x}_{ko} - \dot{x}_o\left(\frac{M_k}{(m_k+M_k)}\right)\right]t(\theta_k) + \left(\frac{M_k}{(m_k+M_k)}\right)x(\theta_k) \quad (22)$$

In order to find $x_k(\theta_k)$ explicitly, we need to find $t(\theta_k)$ and $x(\theta_k)$ and substitute these expressions into equation (22). Using equations (1), (2), (9), (10) and (11), we can show that

$$\ddot{\vec{x}} = -\left(\frac{G(m_k+M_k)}{x^2}\right)\hat{e}_x$$

The above equation can be represented in its scalar form as

$$\ddot{x} - x\dot{\theta}^2 = -\left(\frac{G(m_k+M_k)}{x^2}\right) \quad (23)$$

$$x\ddot{\theta} + 2\dot{x}\dot{\theta} = 0 \quad (24)$$

where $\dot{\theta}$ is the rotation angle of vector $\vec{x}$, the same as the rotation angle of vector $\vec{x}_{M_k}$. The above two equations can be solved to give

$$x(\theta) = \frac{1}{\left[c_1\cos\theta + c_2\sin\theta + \frac{G(m_k+M_k)}{x_o^4 \dot{\theta}_o^2}\right]} \quad (25)$$

where $c_1$ and $c_2$ are constants of integration which can be determined by incorporation of the initial conditions. Now since $\theta = \theta_{M_k} \simeq \theta_k - \pi$, we can write $x(\theta) \simeq x(\theta_k - \pi)$ and show that

$$x(\theta_k) = \cfrac{1}{\left[\cfrac{G(m_k + M_k)}{x_o^4 \dot{\theta}_o^2} - c_1 \cos\theta_k - c_2 \sin\theta_k\right]}$$

where $\dot{\theta}_o = \dot{\theta}(0)$. This can be simplified and written as

$$x(\theta_k) = \frac{1}{\left[k_1 \cos(\theta_k - \phi_k) + k_2\right]} \tag{26}$$

where $k_1 = \pm\sqrt{c_1^2 + c_2^2}$, $k_2 = \dfrac{G(m_k + M_k)}{x_o^4 \dot{\theta}_o^2}$ and $\phi_k = \tan^{-1}\left(\dfrac{c_2}{c_1}\right)$. Having found $x(\theta_k)$, what remains to be done is to solve for $t(\theta_k)$, which can be shown to satisfy

$$t(\theta_k) = \frac{1}{x_o^2 \dot{\theta}_o} \int_{\theta_{ko}}^{\theta_k} x^2(\theta_k) d\theta_k + t_o \tag{27}$$

where $\theta_{ko} = \theta_k(0)$. A substitution of (26) in (27) yields

$$t(\theta_k) = \frac{1}{x_o^2 \dot{\theta}_o} \int_{\theta_{ko}}^{\theta_k} \left[\frac{1}{\left[k_1 \cos(\theta_k - \phi_k) + k_2\right]}\right]^2 d\theta_k + t_o \tag{28}$$

where $\dot{\theta}_o = \dot{\theta}(0)$. Evaluation of the integral gives

$$t(\theta_k) = \left(\frac{2}{x_o^2 \dot{\theta}_o}\right)\left(\frac{k_1 + k_2}{k_2 - k_1}\right)\left[(k_1 + k_2)(k_2 - k_1)\right]^{-\frac{1}{2}} \begin{bmatrix} -k_1 \tan^2(0.5\phi_k - 0.5\theta_k) + k_1 \\ +k_2 + k_2 \tan^2(0.5\phi_k - 0.5\theta_k) \end{bmatrix}^{-1}$$

$$\left\{\begin{array}{l} k_1((k_1 + k_2)(k_2 - k_1))^{\frac{1}{2}} \tan(0.5\phi_k - 0.5\theta_k) + k_1 \tan^{-1}\left[\dfrac{(k_2 - k_1)\tan(0.5\phi_k - 0.5\theta_k)}{((k_1 + k_2)(k_2 - k_1))^{\frac{1}{2}}}\right] \\ \\ + k_2 \tan^2(0.5\phi_k - 0.5\theta_k) - k_1 k_2 \tan^{-1}\left[\dfrac{(k_2 - k_1)\tan(0.5\phi_k - 0.5\theta_k)}{((k_1 + k_2)(k_2 - k_1))^{\frac{1}{2}}}\right] \\ \\ - k_2^2 \tan^2(0.5\phi_k - 0.5\theta_k) \tan^{-1}\left[\dfrac{(k_2 - k_1)\tan(0.5\phi_k - 0.5\theta_k)}{((k_1 + k_2)(k_2 - k_1))^{\frac{1}{2}}}\right] \\ \\ - k_2^2 \tan^{-1}\left[\dfrac{(k_2 - k_1)\tan(0.5\phi_k - 0.5\theta_k)}{((k_1 + k_2)(k_2 - k_1))^{\frac{1}{2}}}\right] \end{array}\right\}$$

$$-\left(\frac{2}{x_o^2 \dot{\theta}_o}\right)\left(\frac{k_1 + k_2}{k_2 - k_1}\right)\left[(k_1 + k_2)(k_2 - k_1)\right]^{-\frac{1}{2}} \begin{bmatrix} -k_1 \tan^2(0.5\phi_k - 0.5\theta_{ko}) + k_1 \\ +k_2 + k_2 \tan^2(0.5\phi_k - 0.5\theta_{ko}) \end{bmatrix}^{-1}$$

$$\left\{\begin{array}{c} k_1((k_1+k_2)(k_2-k_1))^{\frac{1}{2}} \tan(0.5\phi_k - 0.5\theta_{ko}) + k_1 \tan^{-1}\left[\dfrac{(k_2-k_1)\tan(0.5\phi_k - 0.5\theta_{ko})}{((k_1+k_2)(k_2-k_1))^{\frac{1}{2}}}\right] \\ +k_2\tan^2(0.5\phi_k - 0.5\theta_{ko}) - k_1 k_2 \tan^{-1}\left[\dfrac{(k_2-k_1)\tan(0.5\phi_k - 0.5\theta_{ko})}{((k_1+k_2)(k_2-k_1))^{\frac{1}{2}}}\right] \\ -k_2^2 tan^2(0.5\phi_k - 0.5\theta_{ko}) tan^{-1}\left[\dfrac{(k_2-k_1)\tan(0.5\phi_k - 0.5\theta_{ko})}{((k_1+k_2)(k_2-k_1))^{\frac{1}{2}}}\right] \\ -k_2^2 tan^{-1}\left[\dfrac{(k_2-k_1)tan(0.5\phi_k - 0.5\theta_{ko})}{((k_1+k_2)(k_2-k_1))^{\frac{1}{2}}}\right] \end{array}\right\}$$

$$+ t_o \tag{29}$$

Substitution of (26) and (29) into equation (22) would then yield the following expression for $x_k(\theta_k)$.

$$x_k(\theta_k) = x_{ko} - \left(\dfrac{M_k}{m_k + M_k}\right)x_o + \left[\left(\dfrac{M_k}{m_k+M_k}\right)\dot{x}_o - \dot{x}_{ko}\right]t_o + \left(\dfrac{M_k}{m_k+M_k}\right)$$

$$\left[\dfrac{1}{k_1\cos(\theta_k - \phi_k) + k_2}\right] + \left\{\dot{x}_{ko} - \left(\dfrac{M_k}{m_k+M_k}\right)\dot{x}_o\right\} *$$

$$\left\{\left(\dfrac{2}{x_o^2 \dot{\theta}_{ko}}\right)\left(\dfrac{k_1+k_2}{k_2-k_1}\right)[(k_1+k_2)(k_2-k_1)]^{-\frac{1}{2}}\begin{bmatrix}-k_1 tan^2(0.5\phi_k - 0.5\theta_k) + k_1 \\ +k_2 + k_2\tan^2(0.5\phi_k - 0.5\theta_k)\end{bmatrix}\right.$$

$$\left\{\begin{array}{c} k_1((k_1+k_2)(k_2-k_1))^{\frac{1}{2}}\tan(0.5\phi_k - 0.5\theta_k) + k_1\tan^{-1}\left[\dfrac{(k_2-k_1)\tan(0.5\phi_k-0.5\theta_k)}{((k_1+k_2)(k_2-k_1))^{\frac{1}{2}}}\right] \\ +k_2\tan^2(0.5\phi_k - 0.5\theta_k) - k_1k_2\tan^{-1}\left[\dfrac{(k_2-k_1)\tan(0.5\phi_k - 0.5\theta_k)}{((k_1+k_2)(k_2-k_1))^{\frac{1}{2}}}\right] \\ -k_2^2 tan^2(0.5\phi_k - 0.5\theta_k) tan^{-1}\left[\dfrac{(k_2-k_1)\tan(0.5\phi_k - 0.5\theta_k)}{((k_1+k_2)(k_2-k_1))^{\frac{1}{2}}}\right] \\ -k_2^2 tan^{-1}\left[\dfrac{(k_2-k_1)tan(0.5\phi_k - 0.5\theta_k)}{((k_1+k_2)(k_2-k_1))^{\frac{1}{2}}}\right] \end{array}\right\}$$

$$\left. -\left(\dfrac{2}{x_o^2 \dot{\theta}_{ko}}\right)\left(\dfrac{k_1+k_2}{k_2-k_1}\right)[(k_1+k_2)(k_2-k_1)]^{-\frac{1}{2}}\begin{bmatrix}-k_1 tan^2(0.5\phi_k - 0.5\theta_{ko}) + k_1 \\ +k_2 + k_2\tan^2(0.5\phi_k - 0.5\theta_{ko})\end{bmatrix}\right.$$

$$\left\{\begin{array}{l} k_1((k_1+k_2)(k_2-k_1))^{\frac{1}{2}}\tan(0.5\phi_k-0.5\theta_{ko}) + k_1\tan^{-1}\left[\dfrac{(k_2-k_1)\tan(0.5\phi_k-0.5\theta_{ko})}{((k_1+k_2)(k_2-k_1))^{\frac{1}{2}}}\right] \\[6pt] +k_2\tan^2(0.5\phi_k-0.5\theta_{ko}) - k_1k_2\tan^{-1}\left[\dfrac{(k_2-k_1)\tan(0.5\phi_k-0.5\theta_{ko})}{((k_1+k_2)(k_2-k_1))^{\frac{1}{2}}}\right] \\[6pt] -k_2^2\tan^2(0.5\phi_k-0.5\theta_{ko})\tan^{-1}\left[\dfrac{(k_2-k_1)\tan(0.5\phi_k-0.5\theta_{ko})}{((k_1+k_2)(k_2-k_1))^{\frac{1}{2}}}\right] \\[6pt] -k_2^2\tan^{-1}\left[\dfrac{(k_2-k_1)\tan(0.5\phi_k-0.5\theta_{ko})}{((k_1+k_2)(k_2-k_1))^{\frac{1}{2}}}\right] \end{array}\right\} + t_0 \right\} \quad (30)$$

Note that the result presented above is for the generalized case, and therefore it follows that the above result holds valid for the *kth* body of mass $m_k$ $\forall$ $1 \leq k \leq N$.

## Summary:

A set of N gravitating masses was considered and a different procedure was developed to find expressions defining the executed trajectories. It was shown that approximate solutions can be found in closed form, in spite of the fact that exact solutions to such a problem do not exist. We assumed the N body system to be composed of masses having spherically symmetric distributions, small angular velocities (< 1 rad/s) and bounded position vectors. N number of two body motion analogues were then used to approximately replicate N body interaction. The solutions approximately describe the trajectories associated with N body motion in free space.